\RequirePackage{fix-cm}
\documentclass[twocolumn,epjc3]{svjour3}
\smartqed  
\RequirePackage{graphicx}
\RequirePackage{algorithm2e}
\RequirePackage[colorlinks,citecolor=blue,urlcolor=blue,linkcolor=blue]{hyperref}
\usepackage[switch]{lineno}
\usepackage{amssymb}
\usepackage{textcomp}
\usepackage{amsmath}
\journalname{Eur. Phys. J. C}

\begin{document}

\title{The ARTI Framework: Cosmic Rays Atmospheric Background Simulations}

\author{Christian Sarmiento-Cano\thanksref{addr1}
        \and
        Mauricio Suárez-Durán\thanksref{addr2} 
        \and
        Rolando Calderón-Ardila\thanksref{addr3}
        \and
        Adriana Vásquez-Ramírez\thanksref{addr1}
        \and
        Andrei Jaimes-Motta\thanksref{addr1}
        \and
        Luis A. Núñez\thanksref{addr1, addr4}
        \and
        Sergio Dasso\thanksref{addr5,addr6}
        \and
        Iván Sidelnik\thanksref{addr7}
        \and
        Hernán Asorey\thanksref{e1,addr3,addr8,addr9}
        \and
        for the LAGO Collaboration\thanksref{addr10}
}
\thankstext{e1}{e-mail:hernan.asorey@iteda.cnea.gov.ar}
\institute{Escuela de Física, Universidad Industrial de Santander, Bucaramanga, Colombia\label{addr1}
          \and
           Université Libre de Bruxelles (ULB), Brussels, Belgium\label{addr2}
          \and
          Instituto de Tecnologías en Detección y Astropartículas (CNEA, CONICET, UNSAM), Centro Atómico Constituyentes, Buenos Aires, Argentina\label{addr3}
          \and
          Departamento de Física, Universidad de Los Andes, Mérida Venezuela\label{addr4}
          \and
          Departamento de Ciencias de la Atmósfera y los Océanos, Facultad de Ciencias Exactas y Naturales, Grupo LAMP, Universidad de Buenos Aires, Buenos Aires, Argentina\label{addr5}
          \and
          Instituto de Astronomía y Física del Espacio, Universidad de Buenos Aires (CONICET), Buenos Aires, Argentina\label{addr6}
          \and
          Departamento de Física de Neutrones, Centro Atómico Bariloche, (CNEA, CONICET), Bariloche, Argentina\label{addr7}
          \and
          Departamento de Física Médica, Centro Atómico Bariloche, (CNEA, CONICET), Bariloche, Argentina\label{addr8}
          \and
          Unidad de Informática Científica, CIEMAT, Madrid, España\label{addr9}
          \and
          The LAGO Collaboration, see the complete list of authors and institutions at \url{http://lagoproject.net/collab.html}\label{addr10}
}
\date{Received: date / Accepted: date}

\maketitle

\begin{abstract}
ARTI is a complete framework designed to simulate the signals produced by the secondary particles emerging from the interaction of single, multiple, and even from the complete flux of primary cosmic rays with the atmosphere.
These signals are simulated for any particle detector located at any place (latitude, longitude and altitude), including the real-time atmospheric, geomagnetic and detector conditions.
Formulated through a sequence of codes written in C++, Fortran, Bash and Perl, it provides an easy-to-use integration of three different simulation environments: magnetocosmic, CORSIKA and Geant4.
These tools evaluate the geomagnetic field effects on the primary flux and simulate atmospheric showers of cosmic rays and the detectors' response to the secondary flux of particles. In this work, we exhibit the usage of the ARTI framework by calculating the total expected signal flux at eight selected sites of the Latin American Giant Observatory: a cosmic ray Observatory all over Latin America covering a wide range of altitudes, latitudes and geomagnetic rigidities.
ARTI will also calculate the signal flux expected during the sudden occurrence of a gamma-ray burst or the flux of energetic photons originating from steady gamma sources.
It also compares these fluxes with the expected background when they are detected in a single water Cherenkov detector deployed in a high-altitude site.
Furthermore, by using ARTI, it is possible to calculate in a very precise way the expected flux of high-energetic muons and other secondaries at the ground level and to inject them through geological structures for muography applications.

\keywords{Cosmic rays flux \and Space weather \and Gamma Ray Bursts \and Muography}
\end{abstract}

\section{Introduction}\label{sec:introduction}

Except for very particular geographical locations, such as, for example, those active volcanic regions where degassing of $^{222}$Rn into the atmosphere could increase natural radioactivity levels\,\cite{terray2020radon}, the main contribution of the background radiation is due to the continuous arrival of a shower of particles produced during the interaction of primary cosmic rays (CR) with the atmosphere~\cite{kampert2012extensive}.
When an energetic cosmic ray impinges on the Earth's atmosphere, many secondary particles are produced via radiative and decay processes and known as Extensive Air Showers (EAS).
At its maximum development --and depending on the energy $E_p$ of the primary cosmic ray-- the EAS could produce up to $10^{10}$ particles.
The most common technique used in astroparticle physics to study these primary CRs, is the detection of secondary particles at ground level. We can distinguish two important cases: those belonging to a single EAS at the highest primary energy~\cite{pierre2020pierre} and the ones corresponding to the integrated  secondaries due to lower energies~\cite{sidelnik2017lago}.
In this latter case, the precise measurement in the temporal variations of the flux is an important tool in the study of transient astrophysical events, such as the sudden occurrence of gamma-ray bursts~\cite{hurley1994detection,sarmiento2021latin}, space weather phenomena~\cite{Usoskin2008,Asorey2015a} and even geophysical applications like  muography~\cite{Jourden_etal2016,morishima2017discovery,pena2020design}.
Thus, knowing how the primary CRs are modulated by the Earth's magnetic field (EMF) and then reach a given geographical position is critical.
The detailed response of the detector to the secondary particle flux is also essential for a better understanding of the different astroparticle scenarios.

The ARTI toolkit, a set of C++ and Fortran codes, and Bash, Perl and Python scripts, is the result of the efforts of the Latin American Giant Observatory (LAGO) to quantify the effect of the integrated galactic cosmic rays (GCR) flux and many transient astrophysics events at ground level~\cite{Asorey2015a,Asorey2018preliminary}. This toolkit is publicly available at the LAGO GitHub repository~\cite{arti}.
Currently, several computational tools, such as Magnetocosmics (MAGCOS)~\cite{magcos} enable the computation of charged particles' trajectories by using advanced and detailed EMF models.
CORSIKA~\cite{Corsika_1998} or FLUKA~\cite{Fluka_2005} can be used to obtain detailed simulations of extensive air showers in the atmosphere.
Finally, the detector response is simulated by Geant4~\cite{Agostinelli2003}, a widespread tool used to simulate the interaction of radiation with matter developed by the high energy physics researchers.
However, while these Monte Carlo tools are being extensively developed and tested within the Astroparticle Physics community, they cannot integrate secondary background flux produced by the total flux of primary cosmic rays, or to concatenate all the intermediate results.
All these processes have to be done ``by hand'' every time a detection location needs to be characterized or evaluate the impact of the changing atmospheric or geomagnetic conditions.
In the end, we are performing all these operations manually, which is a hard and tedious jobs. Thus, we developed ARTI, to carry out these tasks in a semi-autonomous way~\cite{arti}. The framework calculates and analyses effortlessly the total background flux of secondaries and the corresponding detector signals produced by the atmospheric response to the primary flux of GCR\@.
This analysis can be implemented for single or small arrays of Water Cherenkov Detector (WCD) located at any geographic site, and under realistic and time-evolving atmospheric, geomagnetic and detector conditions.

\begin{figure}[ht] \centering
	\includegraphics[scale=0.8]{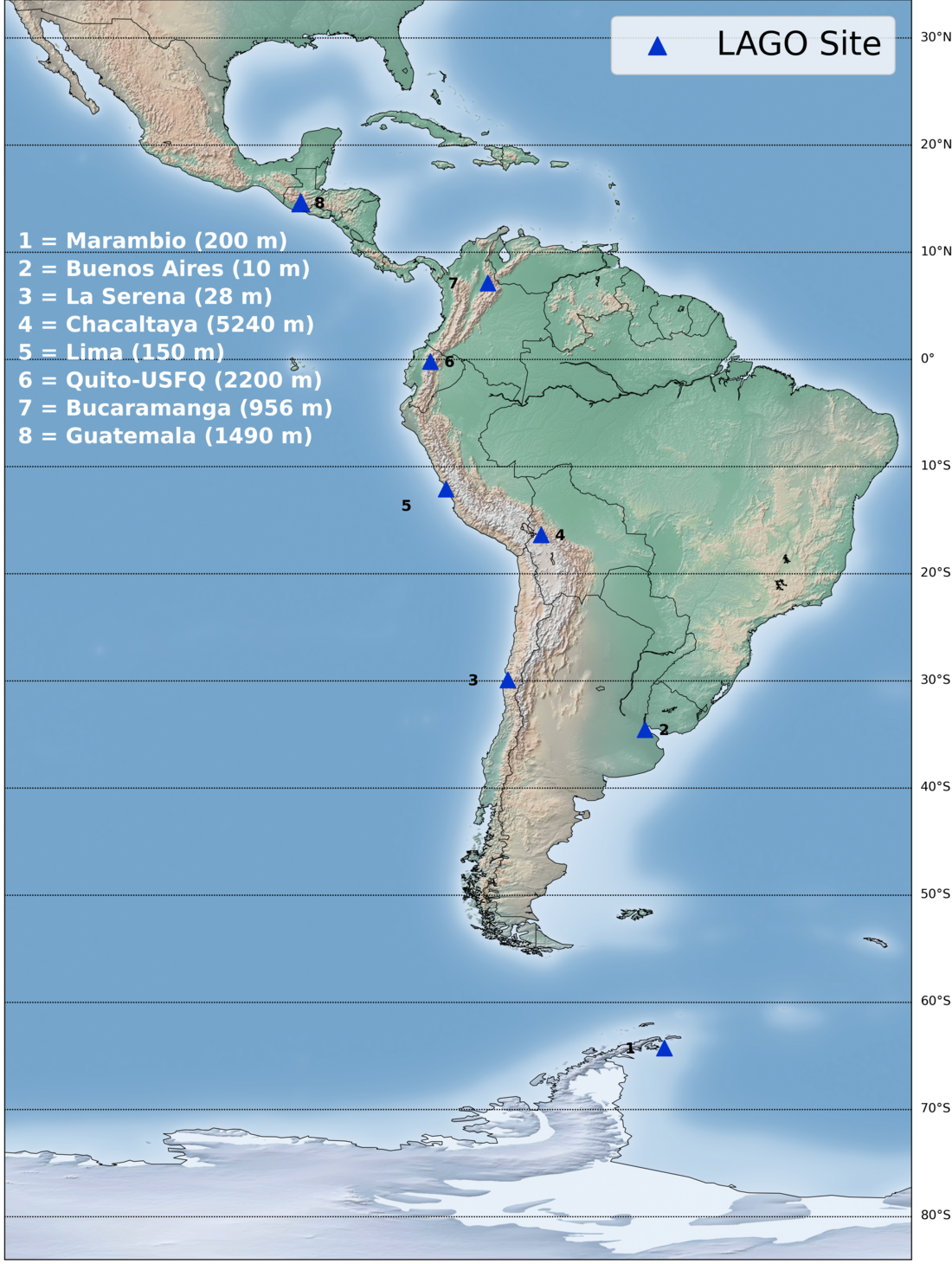}
	\caption{
		In this work we studied these eight representative sites of the Latin American Giant Observatory, covering a wide range of altitudes and local geomagnetic rigidity cut-offs, from south to north: Marambio, Antarctica; Buenos Aires, Argentina; La Serena, Chile; Chacaltaya, Bolivia; Lima, Perú; Quito, Ecuador; Bucaramanga, Colombia; and Guatemala, Guatemala. Detailed information about these sites can be seen in Table~\ref{tab:sites}.
	}\label{fig:figlagomap}
\end{figure}

As an example of the ARTI's functionality, we conduct a comparative study of the response of a LAGO WCD to the total flux of secondary particles at eight LAGO sites in Latin America.
We select these sites --shown in figure~\ref{fig:figlagomap}-- to emphasize the differences in their altitudes, local atmospheric conditions and geographic and geomagnetic coordinates.

The outline of this document is as follows: in section~\ref{sec:artimethod} we describe the computational architecture of ARTI\@.
Next, section~\ref{sec:astrobackground} introduces the method implemented for estimating the nominal background flux of secondaries at the detector level.
Then, in section~\ref{sec:geomagneticfield}, we explain the methodology used to implement the correction due to the EMF condition in the selected LAGO sites~\cite{Asorey2018preliminary}.
The Geant4 models and the signals expected in a standard LAGO WCD are shown in section~\ref{sec:secwcd}, where we use the same type of detector to simplify the direct comparison of the results in different sites.
We discuss the final remarks and future perspectives of this work in section~\ref{sec:discussions}.
Finally, in~\ref{sec:pseudocode} and~\ref{sec:tutorial} we sketch the main parts of the framework and include a brief description of some main options to be used in different types of cosmic ray simulations.

\section{ARTI framework}\label{sec:artimethod}

With ARTI it is possible to calculate the expected signal flux at any WCD and, depending on the user's initial configuration, ARTI calls the corresponding modules to interact with CORSIKA or Geant4.

The ARTI approach is in three stages:
\begin{enumerate}
    \item site characteristics, primary spectrum calculations and EAS developments;
    \item analysis of the secondary particles at the ground and for geomagnetic corrections;
    \item and for detector simulation.
\end{enumerate}

These three stages are sequentially accessed: the physical results from one stage are used as the input for the next stage.
The files resulting from each stage are preserved and curated as LAGO datasets~\cite{rubiomontero2021novel,dmp}. In this section we present the overall simulation sequence, while some main steps and the intermediate physical results are detailed in the following sections.

\begin{figure*}[ht]
\centering
	\includegraphics[scale=0.625]{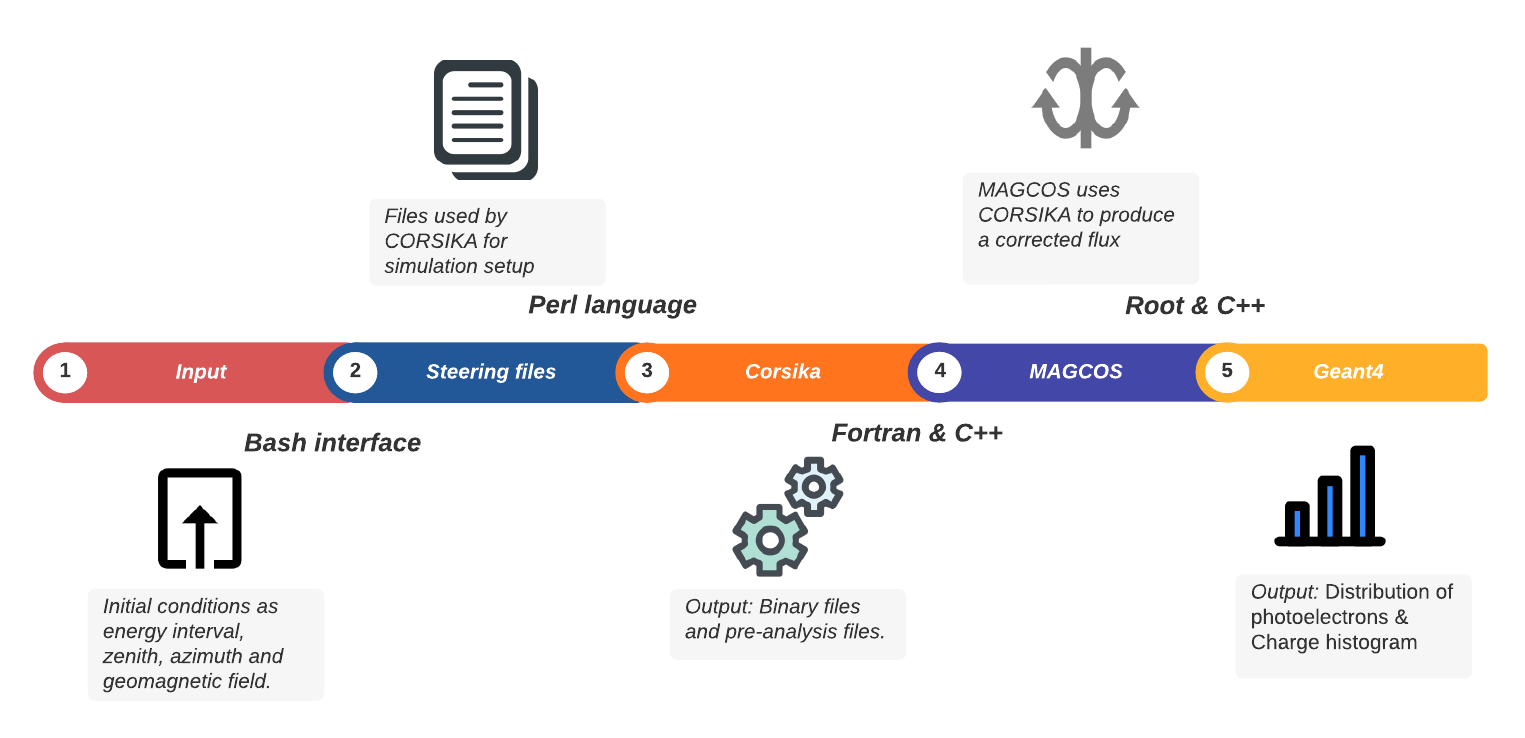}
	\caption{As the simulation evolves, the different ARTI modules call CORSIKA, MAGCOS, or Geant4 to accurately calculate the signal flux at the detector expected by the integrated cosmic ray flux. The simulation moves forward as a sequence: the results from one steps are used as inputs for the next one.}
\label{fig:fluxdiagram}
\end{figure*}

The ARTI framework consists of C++ and Fortran codes; Perl and various bash scripts which provides the interaction between the user and the different software (for details, please see~\ref{sec:pseudocode}).
This allows a straightforward approach for the user and a smooth interaction between ARTI and CORSIKA, MAGCOS and Geant4, as shown in figure~\ref{fig:fluxdiagram}.

Different tracks depend on the type of calculations the user would need. We will detail the necessary steps in calculating the expected signal flux at a particular WCD\@.
We refer the reader to the ARTI website for further explanations and examples~\cite{arti}.
The main script for this calculation is {\texttt{do\_sims.sh}}.
All ARTI scripts can be configured using options and modifiers from the command line (cli).
ARTI versions could change, but all the ARTI scripts and codes have an internal help modifier, ({\texttt{-?}}).
The currently available set of options ({\texttt{do\_sims.sh}}) is available in~\ref{sec:tutorial}.
Here we describe the main simulation sequence with their corresponding options and modifiers for the initial configuration, i.e.:
\begin{description}
    \item[{\texttt{-t}}], defines the time that will be used for the integration of the primary flux.
    Typical values range from $3600$\,s to $86400$\,s per run.
    \item[{\texttt{-s}}], specifies the geographic from a list of predefined sites around the World.
    Each site has an altitude above sea level, atmospheric parameters and local values of the EMF, as described in the LAGO Data Management Plan (DMP)~\cite{rubiomontero2021novel,dmp}.
    Otherwise, the corresponding atmosphere model (\texttt{-c}), altitude (\texttt{-k}) or the $B_x$ (north) and $B_z$ (vertical) components of the magnetic field could be defined (\texttt{-o} and \texttt{-q} respectively). ARTI requires them to calculate the local EMF corrections at a later stage (see section~\ref{sec:geomagneticfield}) and by CORSIKA to take into account the geomagnetic effects on the primary and secondary charged particles propagation in the atmosphere.
    \item[{\texttt{-c}}] defines which type of atmospheric model will be used from these possibilities: a) atmospheres from the MODTRAN atmospheric model~\cite{Kneizys1996modtran};
    b) atmospheres characterised by the 20 parameters following the well known Linsley's atmospheric model;
    or c) a realtime extraction of the instantaneous atmospheric profiles as compiled from the Global Data Assimilation System (GDAS)~\cite{Grisales2021impact}.
    \item[{\texttt{-u}}] is the ORCID identifier of the user, for further reference of the results dataset~\cite{dmp}.
    \item[{\texttt{-v}}] represent the CORSIKA version to be used by ARTI. The Current default is version CORSIKA 7.7402.
    \item[{\texttt{-y}}] is required to calculate the flux by considering a volumetric detector (such as a WCD), instead of a flat detector (such as a scintillator panel or a resistive plate chamber).
    \item[{\texttt{-x}}] is the default for the non-selected options, just use the internal ARTI predefined values (energy range, arrival directions, etc.)
\end{description}

A typical call for the initial script could be:
\begin{verbatim}
$ do_sims.sh -w path_to/corsika/run/ -p fluxBGA
   -s bga -v 76500 -t 3600 -u user -y -x
\end{verbatim}

ARTI will calculate the one-hour ($3600$\,s) integrated flux of secondaries for a volumetric detector (\texttt{-y}) located at ground level in the LAGO site of {\texttt{bga}} (Bucaramanga, Colombia, 965 m a.s.l.)\footnote{When possible, LAGO sites are internally known by the corresponding three or four letters IATA code of the local airport.}, following the calculations described in section~\ref{sec:astrobackground}, selecting the v7.6500 version of CORSIKA and the ARTI defaults ({\texttt{-x}}) for the non-selected options (see the complete list in~\ref{sec:tutorial}). All the files will be located in the subdirectory {\texttt{fluxBGA}} created in the CORSIKA executable's main path.

By default, ARTI divides the total flux of particles into 60 processes and produce 12 bash scripts to distribute the simulation load in a computer cluster.
As described in section~\ref{sec:astrobackground}, each process is determined by considering the flux of an individual particle that constitutes the primary background, by injected primaries from proton to iron.
The default number of processes can also be modified by the user (\texttt{-j} option).
Depending on the computing environment, it will produce an additional bash script to perform the automatic analysis of the results. It will automatically launch the CORSIKA stage simulations in, e.g., the corresponding user-accessible queues in high performance computing (HPC) clusters operating with the SLURM workload manager, or in cloud-based environments with docker containers~\cite{rubiomontero2021novel,rubiomontero2021eosc}.
At the end of this first stage, if the default number of processes was used, ARTI will produce 60 CORSIKA steering files (\texttt{*.input}), 60 bzip2-compressed unformatted binary fortran files containing all the simulated EAS information (\texttt{DAT*.bz2}), and 60 bzip2-compressed files containing the output produced by CORSIKA during the simulation run (\texttt{*.lst.bz2}).
According to the LAGO DMP, all the files resulting from this first simulation stage are considered a data catalogue and are collectively known as the {\texttt{S0}} datasets.

The {\texttt{S0}} produces catalogues that can be read and analyzed with the ARTI set of tools which extracts the secondary particle flux information, i.e.\ the secondary particle ID, the particle momentum, the particle position relative to the primary core, and additional information related with its primary progenitor.
The analysis stage can be performed automatically during the runtime or manually by calling the {\texttt{do\_showers.sh}} script. The typical result of this script is several bzip2-compressed ASCII files containing the information of the injected primaries (\texttt{*.pri.bz2}), the resulting secondaries (\texttt{*.sec.bz2}), and four additional files: 
\begin{itemize}
    \item \texttt{.shw.bz2} file containing the complete list of secondary particles at ground level and information about their corresponding progenitors;

    \item \texttt{.hst} file containing the energy distribution of the secondary particles by particle type;
    
    \item \texttt{.dse} and \texttt{.dst} files containing, respectively, the lateral distribution of the deposited energy and lateral distribution of the number of particles at the ground, respectively.
\end{itemize}

Depending on the computing environment, there are different types of parallelized analyses available that can be selected from cli. The recommended number of parallel processes are also automatically selected from the local node's total number of available processors.
If the files to be analyzed are stored on cloud servers, they are first transferred to the local cluster before processing.
For example, a standard manual call for running a parallelized analysis of a {\texttt{S0}} catalogue stored in a cloud storage server, corresponding to the flux during one day at a site with altitude of $2400$\,m a.s.l., could be:

\begin{verbatim}
$ do_showers.sh -o path_to/cloud_server/fluxBGA
-k 2400 -t 86400 -g site.geo 3 -l
\end{verbatim}

In this case, the default ARTI option implements a parallel analysis using half of the available processors at each node.
During the analysis the EMF is taken into account, and the local directional rigidities (see below) are read from the third column of the {\texttt{site.geo}} file.
While this can be called by the user at any moment for, e.g., to reprocess the same {\texttt{S0}} catalogue for different EMF conditions, this stage is usually performed automatically by ARTI using the predefined default options without need of the user interaction.
Please refer to the ARTI web site~\cite{arti} and~\ref{sec:tutorial} for the detailed information about these scripts.

The predefined values for the local EMF are calculated and regularly updated~\footnote{The values for the magnetic field components can be automatically updated during runtime by the using the option \texttt{-g}, or forcing the values of $B_x$ and $B_z$ at launch time by using the corresponding options.} following the latest available version of the International Geomagnetic Field Reference (IGRF)\footnote{Currently  IGRF13-2019~\cite{Alken2021international}}.
Notice that, at this stage, only the secular --very long term variations-- of the EMF were taken into account. However, at this phase, we have not considered the effects due to the penumbra of the magnetic field. They are usually described as a system of forbidden and allowed trajectories within an intermediate range of local geomagnetic cut-off rigidities~\cite{Bobik2001geomagnetic}.
The Earth's magnetic field is continuously modulated and disturbed by the Solar activity.
Therefore, as explained in section~\ref{sec:geomagneticfield}, we provide an additional script to include both effects automatically.
To perform this analysis in ARTI one must type:
\begin{verbatim}
$ STATIC_MAGNETOCOSMICS base.g4mac
\end{verbatim}
creating a file with the corresponding values for the directional upper and lower limits of the local rigidities, at the site on a given date. This also is helpful for filtering and eliminating those secondaries in a forbidden primary within a \texttt{.shw} file. This method allows the re-use of the same {\texttt{S0}} catalogue for different instantaneous disturbances of the EMF\@.
All the resulting files from the analysis stage are also considered a new LAGO data catalogue and are collectively identified as a {\texttt{S1}} dataset.

Finally, a Geant4-based code in ARTI calculates the WCD response to the secondary particles arriving at the detector.
As explained in section~\ref{sec:secwcd}, ARTI injects every secondary particle within the detector volume with its initial momentum but with a random position above the WCD\@.
The Geant4 simulation includes different parts that make up the detector, such as: 
\begin{itemize}
    \item the water container material, thickness and geometry;
    \item the reflectivity and diffusivity of the container's inner coating made of Tyvek{\textregistered}; 
    \item the water absorption curves in the optical and UV wavelength range; the geometry and wavelength sensitivity of the photomultiplier tube (PMT); 
    \item and the emulated response of the onboard LAGO electronic system.
\end{itemize}

The user can vary all these characteristics so as to incorporate the ones corresponding to the WCD to be simulated.

Depending on the environment, it could run automatically during execution or will be called by the user once the previous stages are completed, by typing:
\begin{verbatim}
$ wcd -m input.in
\end{verbatim}

Once this stage ends, we have a {\texttt{.root}} file containing the expected charge histograms at the detector, i.e., the histograms of the time integral of the individual pulses produced by each secondary particle impinging in the detector.
As in the first two stages, ARTI identifies the resulting files from this last stage as the {\texttt{S2}} LAGO data catalogue.

The ARTI framework relies on the previous installation and compilation of ROOT~\cite{Brun1997root}, CORSIKA and Geant4.
The current version of ARTI (v1r9) works with CORSIKA v7.7402, ROOT v6.20.08, Geant4.10.03 and Magnetocosmics v2.0 versions.
Anyway, backward compatibility within previous minor versions and patches of the external dependencies is assured.

Once these tools are correctly installed, ARTI can be easily cloned from the LAGO GitHub repository~\cite{arti} and automatically installed by running the provided installation script. As usual, ARTI performance will depend on the system and the installation options. For a typical installation in a modern HPC cluster, i.e.\ a cloud virtual cluster based on a Slurm Workload Manager job scheduler with one virtual master (v-master) and 10 v-nodes with 8 Intel Xeon Core E7 processors and 250 GB of shared memory~\cite{rubiomontero2021eosc}. All the computations are completed almost in real time. To compute the signals expected in a WCD deployed in a low-latitude site during 1 day of total primary flux ($\lesssim 10^9$ primaries) requires $\simeq 28$\,hours ($\simeq 2.2$\,kCPU$\cdot$hours) and produce $\sim 30$\,GB of compressed ({\texttt{S0}}+{\texttt{S1}}+{\texttt{S2}}) datasets.

The following sections describe the physical process and the results obtained for each of the three stages.

\section{Cosmic background radiation at ground level}\label{sec:astrobackground}

We model the GCR primary flux ($\Phi$) as an isotropic flux impinging the Earth's atmosphere at an altitude of $112$\,km a.s.l.~\cite{AguilarAMS-01}, where $\Phi$ is defined as
\begin{equation}
    \centering
    \Phi(E_p, Z, A, \Omega) \simeq j_0(Z,A) \left(\frac{E_p}{E_0}\right)^{\alpha(E_p,Z,A)}\, ,\label{eq:primaryflux}
\end{equation}
with $E_p$ the energy of the primary particle, $\alpha(E_p,Z,A)$ the spectral index for each type of injected particle and which can be considered constant, i.e. $\alpha\equiv\alpha(Z,A)$, in the energy range of interest in this work, from a few GeV to $10^{6}$\,GeV ($1$\,PeV).
Each kind of GCR considered is individualized by its mass number ($A$) and atomic number ($Z$) and $j_0(Z,A)$ is the measured flux in the top of the atmosphere at the reference energy $E_0=10^3$\,GeV. See~\cite{Asorey2018preliminary} and references therein for a detailed description of this method.

To calculate the expected number of primaries, we integrate $\Phi$ for each primary type from proton to iron (i.e., $1\leq Z \leq 26$).
Please notice that the integrated flux is simulated with the corresponding distribution for each type of particle~\cite{Asorey2015a}.
We use time integration values typically ranging from one hour to one day of flux impinging over an area of $1$\,m$^2$ and isotropically distributed in $0\leq\theta\leq\pi/2$ and $-\pi\leq\phi\leq\pi$ where $\theta$ and $\phi$ are the zenith and azimuth angles respectively.

If the rigidity cutoff option is considered (option {\texttt{-b}}), then the minimum energy $E_{\min}(Z)$ is selected for each type of primary as $E_{\min}(Z) = R_0 \times Z$, where $R_0$ is lower than the minimum value of the local directional rigidity cut-off $R_{\min}$, i.e., $R_0 < R_{\min}$.
Otherwise, a minimum value of $0.1$\,GeV for the kinetic energy is used for all the injected primaries, and so, in natural units, $E_{\min}=m(Z,A) + 0.1$\,GeV, where $m(A,Z)$ is the mass of the injected primary.
In all the cases, the upper limit of the energy range is fixed to $E_{\max} = 1$\,PeV, as for $E_p > E_{\max}$ a decrease in the GCR flux is observed for all species (the so-called knee of the cosmic ray spectrum, see e.g.~\cite{Letessierselvon2011ultrahigh}), and at this energy the flux can be considered so low that it does not affect the background calculations\footnote{This is ARTI's default behavior, but it can be modified at launch time by using the {\texttt{-r}}, {\texttt{-l}}, and {\texttt{-b}} options. See~\ref{sec:tutorial} and~\cite{arti} for further information.}.

Given the Poissonian behavior of the primary flux, all the computations can be easily done in parallel  as EAS do not interact with each other.
As for energies below the knee, protons then helium dominate the GCR flux. To rest of the nuclei corresponds to a small fraction of the total flux.
Once we calculate the total number of particles to be injected for each primary $(A,Z)$, they are grouped looking forward uniforming all computing aspects. For the physical aspects to be considered, the computing time not only depends on the total number of primaries but also on the energy ranges and the spectral index $\alpha(A,Z)$: low energy particles are much more abundant, but take much less time to compute.
As described below, the site altitude effect is not trivial: at high altitudes sites, the cascades reach the ground during early stages of their development, but at the same time, the total number of secondaries that need to be tracked is much larger than at later stages of the EAS development.
When the site altitude is close to the EAS $X_{\max}$ (the atmospheric depth where its reaches its maximum development), the total number of air shower increases.

As mentioned in section~\ref{sec:artimethod}, ARTI currently supports different types of cluster architectures and distributed computing  solutions, such as those based on grid and federated or public clouds implementations~\cite{rubiomontero2021novel}.
So, from the computing point of view, it is relatively common than the number of available CPUs per nodes at HPC infrastructures was a power of 2.
Additionally, if the binary files result too large, they can quickly fill the available local storage before processing, with the corresponding loss of all the calculations on this node.
Thus, after performing careful studies, the grouping or splitting of primary species results as follow: a) the total flux of protons is splitted into $32$ separated processes;
b) the total flux of Helium is splitted in $8$ separated processes;
and c) the non-dominant components are grouped following this schema: ($^{12}$C-$^{16}$O-$^{7}$Li-$^{24}$Mg); ($^{11}$B-$^{28}$Si-$^{14}$N-$^{20}$Ne); ($^{56}$Fe-$^{9}$Be-$^{32}$S-$^{27}$Al); ($^{23}$Na-$^{40}$Ca-$^{19}$F-$^{52}$Cr); ($^{40}$Ar-$^{48}$Ti-$^{55}$Mn-$^{39}$K); and ($^{51}$V-$^{31}$P-$^{35}$Cl-$^{45}$Sc).
Future releases of ARTI will be optimized by mixing protons and helium nuclei with the non-dominant species.

The atmospheric profile is another critical factor in producing secondary particles at ground level. Thus, we need to set up the model for each site's geographical location with at least the season profiles of MODTRAN~\cite{Kneizys1996modtran}, or the exact time instead of the atmosphere profile extracted from GDAS~\cite{Grisales2021impact}.
Depending on the atmospheric model selected by the user at launch time, ARTI will automatically handle all the internal CORSIKA steering keywords needed to run the EAS simulation.

Once the user sets all the primary simulation parameters (say, integration time and area;
energy and solid angle ranges;
site altitude, atmospheric model and EMF coordinates;
atmospheric model, etc), ARTI calculates the number of particles of each kind that will be injected by integrating equation~\eqref{eq:primaryflux}, split or group at the primaries following the above described schema; creating the corresponding CORSIKA {\texttt{.input}} data files\footnote{A specially formatted ASCII file containing all the parameters needed to run the corresponding CORSIKA simulation.}, and the scripts that will run the EAS simulation processes.
These scripts are automatically adjusted depending on the computing environment where ARTI is running, and they are eventually launched or they are queued in the cluster's partition.
After starting the first stage of ARTI calls the CORSIKA executable to simulate the cascades produced during the interaction of the primaries with the atmosphere.

For this example we used CORSIKA version 7.6500 compiled with the following options and modules: QGSJET-II-04~\cite{Ostapchenko2011}; GHEISHA-2002;
EGS4; curved and external atmospheric modules and flux calculations for a volumetric detector.
For the selected eight sites of this example, we used the standard MODTRAN profiles: a tropical profile for Bucaramanga (BGA, Colombia), Ciudad de Guatemala (GUA, Guatemala) and Quito (UIO, Ecuador); a subtropical summer profile for Buenos Aires (EZE, Argentina), Lima (LIM, Perú), La Serena (LSC, Chile) and Chacaltaya (CHA, Bolivia), and the summer antarctic profile for the Marambio Base (SAWB, Antarctica)~\cite{DassoEtal2015,MasiasMarambio}.
The local EMF components $B_x$ and $B_z$ at the site of this example were obtained from the IGRF-12 model~\cite{thebault2015international}.

As the EAS simulation evolves, each secondary particle is tracked down to ground level (the observation level), unless the particle decays or its energy $E_s$ goes below the lowest secondary energy threshold $E_{\mathrm{cut}}$ allowed by CORSIKA\footnote{These thresholds depends on the CORSIKA version and the type of secondary. For the current versions of CORSIKA, these thresholds are E$_s$\,$\geq$\,5\,MeV for $\mu^\pm$s and hadrons (excluding $\pi^0$s); and E$_s$\,$\geq$\,5\,keV for e$^\pm$s, $\pi^0$s and $\gamma$s. While these limits can't be decreased, they can be increased in ARTI by using the option {\texttt{-a}}, as this can be very useful for high energy applications such as muography, where the secondaries needs then to go through, e.g., $500$\,m of rock~\cite{taboada2022meiga}.}.

\begin{figure*}[ht]
\centering
	\includegraphics[scale=0.625]{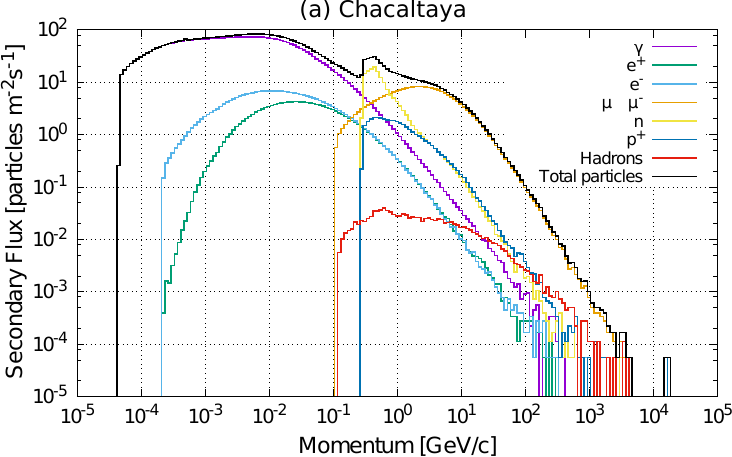}
	\includegraphics[scale=0.625]{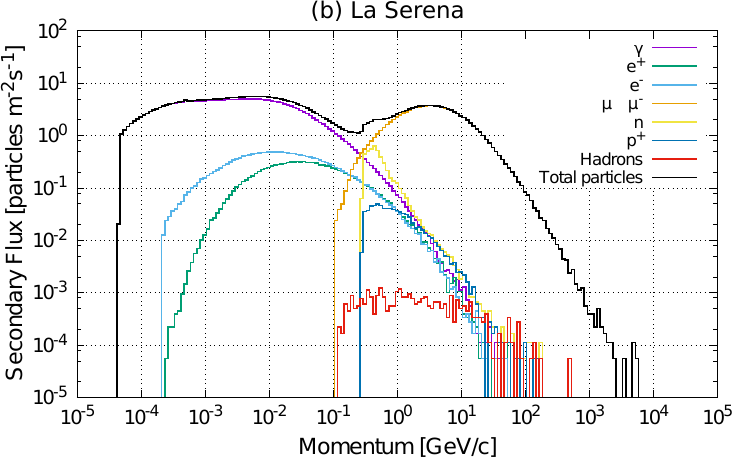} \\
	\includegraphics[scale=0.625]{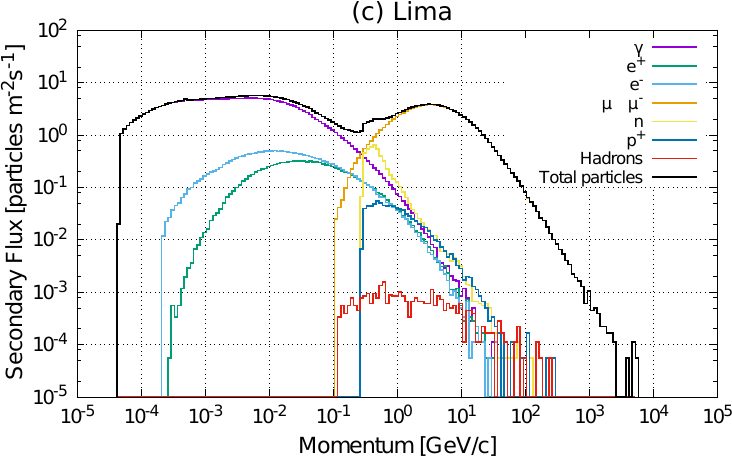}
	\includegraphics[scale=0.625]{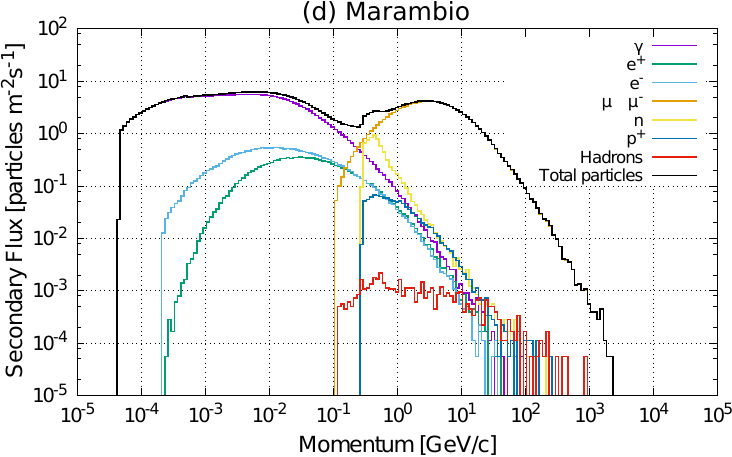}
    \caption{
		Spectra of the secondary particles discriminated by particle type at four LAGO sites: (a) Chacaltaya (CHA), Bolivia ($5240$\,m a.s.l.); (b) La Serena (LSC), Chile ($28$\,m a.s.l.); (c) Lima (LIM), Perú ($150$\,m a.s.l.) and (d) Marambio Base (SAWB), Antarctica ($200$\,m a.s.l.). The comparison of these four plots evidences the atmospheric absorption, reaching up to one order of magnitude (at $\sim$\,10$^{-2}$\,GeV/c) in the total flux of secondaries (black line). See a detailed description of this figure in the text.
    }\label{fig:RelativeFlux}
\end{figure*}

Once the EAS simulations end, the second stage of ARTI is called to obtain the total secondary flux of particles at the ground, $\Xi$.
Again, after the trivial parallelization method we performed in the first stage, we take advantage of the poissonian nature of the cascades and the uniformity of the primary flux to superpose all the resulting secondaries (contained in the {\texttt{.sec}} files) to generate a single file (the so-called \lq\lq{}shower\rq\rq{} {\texttt{.shw}} file) contained all the resulting particles at ground level.
The stochastic nature of the primary flux and the EAS development, allows the possibility to normalize the flux by considering that all the secondaries produced by the primaries impinging in an area $\Delta A$ (typically, $\Delta A=1$\,m$^2$) at the upper atmosphere during the time $t$, will reach the ground level in the same surface $\Delta A$ and in the same time period $t$.
While the assumption on the later is easy to understand, as the time evolution of the cascade is very short ($\mu$s) when compared with the typical of values considered for $t$, the former is not obvious at all, as it is in the basis of the definition of an EAS: the almost simultaneous arrival of secondary particles distributed over a large area at ground.
This distribution is originated by the transverse momentum $p_{\perp}$ acquired by the air shower particles at production and the subsequent scattering in the atmosphere.
As an order of magnitude, the electromagnetic Moliere radius is of $X_m = 9.3$\,g\,cm$^{-2}$, which in air and at sea level corresponds to $R_m=80$\,m~\cite{Zyla2020review}.
However, it is important to recall on the uniformity and isotropy of the flux at this energy ranges.
In average, a secondary particle that miss the target area at ground by, say, $50$\,m to the north, will be compensated by, a quite similar secondary particle produced during the cascade development of an equivalent primary impinging the upper atmosphere at $50$\,m to the south.
This kind of sib-similarity properties of the flux of secondaries at ground is the basis of the ARTI spatial normalization of the flux $\Xi$ of secondaries at ground, and of the flux $\Xi^{\mathrm{D}}$ measured at the detector (see section~\ref{sec:secwcd}).

As an example, in figure~\ref{fig:RelativeFlux} the results obtained for the expected secondary spectra at four of the eight representative sites are shown: Chacaltaya (Bolivia) at $5240$\,m a.s.l., La Serena (Chile) at sea level ($28$\,m a.s.l.), Lima (Perú) at $150$\,m a.s.l.\ and Marambio (Antarctica) at $200$\,m a.s.l.\
There are several features that can be seen in the secondary spectra.
Typically, secondary particles are grouped in three main groups: the electromagnetic component, composed of $e^\pm$ and $\gamma$s; the muonic $\mu^\pm$ component;
and the hadronic component composed by all the barions and mesons that are present in the cascade.
However, in this work we separate neutrons and protons from other hadrons as they can be used as tracers for some cascade mechanisms (protons) or space weather activity (neutrons)~\cite{sidelnik2020simulation}.
As can be seen, while at low secondary momentum $p_s$ the flux is dominated by the electromagnetic component, the high energy flux is totally dominated by muons, and even for this short integration time it is possible to observe some muons that could reach up to several tens of TeV/c.
These muons posses enough energy to traverse hundreds and up to thousands of meters of rock and could be the main source for signals in muography studies~\cite{bonechi2020atmospheric} or background noise at underground laboratories~\cite{perezbertoli2022estimation}.
In general, atmospheric absorption produces the well known decrease in the total flux of secondaries at low altitudes in all components, as can be seen comparing, the two left side panels of Figure~\ref{fig:RelativeFlux}, where a difference of up to an order of magnitude exists between the integrated flux at Chacaltaya (CHA) that is at more than $5,000$ m a.s.l., and Lima at sea level.
Even more, at this stage, the simulation results are so detailed that it is even possible to see a significant increase in the photon flux at the $510-520$\,keV energy bin, corresponding to the $511$\,keV photons produced during pair annihilation in the atmosphere, as it is also reflected in the differences in the flux of electrons and positrons at low $p_s$.

Additionally, the cascade evolution through the atmosphere can be inferred from the relative fraction of components.
For example, at Chacaltaya altitudes the charged-pions-to-muons fraction is larger than at sea level, as most of the muon component is originated from the charged pions decay, a process that typically occurs below $5,000$\,m a.s.l.\ At $\sim$\,3.5$\times10^{-1}$\,GeV/c a comparison between the flux of neutrons at Chacaltaya and La Serena shows that at the highest altitudes the flux in this $p_s$ range is dominated by neutrons instead of muons, while at lower altitudes the effect is opposite.
Since the LAGO detector calibrations are based on muons, it is important to notice also that the prediction for the muon component is larger than the charged electromagnetic component at La Serena, at sea level, while, at the altitude of Chacaltaya, the $e^\pm$ dominates with respect to $\mu^\pm$ due to atmospheric development of the cascades.
This kind of studies are frequently used in LAGO and in other astroparticle observatories to characterise the planned sites for different types of astrophysical studies.

\section{Geomagnetic Field Corrections}\label{sec:geomagneticfield}

The Earth's magnetic field deflects low energy GCRs ($E_p<100$\,GeV) trajectories which are usually parametrized by the magnetic rigidity term (R$_\mathrm{m}$)~\cite{Smart2009fifty,Modzelewska-Alania-2013,masias2016superposed}.
During these events, the geomagnetic field can also be disturbed due to its interaction with the magnetized solar wind plasma, consequently changing the geomagnetic shielding on energetic particles and modifying their trajectories.
Thus, Solar activity could change the primary flux and so, also change the observed flux of secondaries at ground level~\cite{Usoskin2008,Cane2000,Asorey2011}, as  e.g., the well known Forbush decreases (FD) are manifestation of this phenomena.

Several FDs have been registered by different cosmic ray observatories using WCDs~\cite{Asorey2015a,Asorey2011,Angelov2009forbush,Dasso2012scaler,Mostafa2014high}.
In 2013 the LAGO Collaboration developed its LAGO Space Weather (LAGO-SW) program~\cite{Asorey2013,Asorey2015a}, to study the variations in the flux of secondary particles at ground level and their relation to the heliospheric modulation of GCRs.

The EMF effect on the flux $\Xi$ has been included in this work for each of the eight selected LAGO sites, following our filtering method during the second stage of the ARTI framework.
In this method, the local magnetic rigidity cut-off ($R_\mathrm{c}$) defined as,
\begin{equation}\label{eq:renewrm}
	R_{\mathrm{c}} = R_{\mathrm{c}}\left(\mathrm{Lat},\mathrm{Lon},\mathrm{A lt},\mathrm{T},\theta,\phi,P\left(R_{\mathrm{m}}\left(\theta,\mathrm{T}\right)\right)\right),
\end{equation}
is a function of the geographical latitude (Lat), longitude (Lon), altitude above sea level (Alt), primary arrival direction $(\theta,\phi)$ and epoch time $T$, is used to determine whether a secondary particle should reach the ground or not depending on the progenitor primary rigidity, and including a cumulative probability distribution function for the penumbra region, $P\left(R_\mathrm{m}\left(\theta,\mathrm{T}\right)\right)$, as explained in~\cite{Asorey2018preliminary}.
These methods allow us to determine the expected background flux baseline and to evaluate the impact of the changing geomagnetic conditions during, for example, a geomagnetic storm.

Table~\ref{tab:sites} displays some of the main characteristics of the selected LAGO sties and the results for the estimated flux of particles at ground level, including the EMF correction.
It is clear that, as mentioned in section~\ref{sec:astrobackground}, for sites with similar secular values of $R_\mathrm{c}$ the site's altitude is the predominant variable.

\begin{table*}[ht]
  \centering
  \caption{
	  The estimated flux of cosmic background radiation at ground level for eight LAGO sites: Chacaltaya, Quito, Guatemala, Bucaramanga, Marambio Base, Lima, La Serena, and Buenos Aires, characterized by their geographical coordinates, atmospheric profiles and altitude above sea level. The total flux for all secondaries ($\Xi^{\mathrm{All}}$) is presented along with the relative geomagnetic effect over the total flux (GE$^{\mathrm{All}}$[\%]), estimated as the percent difference with respect to the flux $\Xi$ without the EMF corrections.
  }\label{tab:sites}
  \begin{center}
  \begin{tabular}{|c|c|c|c|c|c|c|c|c|}
    \hline
    LAGO & Site & Country & Coordinates & MODTRAN &  Alt & $R_\mathbf{c}$ & $\Xi^{\mathrm{All}}$ & GE$^{\mathrm{All}}$ \\
    site & name & or region & (Lat, Lon)  & profile    & m a.s.l.\  &  GV/c &    m$^{-2}$ s$^{-1}$ &       $\%$                \\
    \hline
    \hline
    CHA & Chacaltaya & Bolivia & (16.35 S, 68.13 W) & subtropical summer & 5240 & 11.6 & 4450 & -18\\
    \hline
    UIO & Quito & Ecuador & (0.20 S, 78.5 W)  & tropical & 2800 & 12.2 & 1260 & -12 \\
    \hline
    GUA & Guatemala & Guatemala & (14.63 N, 90.59 W) & tropical & 1490 & 9.1 & 730 & -4 \\
    \hline
    BGA & Colombia & Colombia & (7.14 N, 73.12 W)  & tropical & 956 & 11.6 & 540 & -6 \\
    \hline
    SAWB & Marambio & Antarctica & (64.24 S, 56.62 W) & antarctic summer & 200 & 2.2 & 430 & -1 \\
    \hline
    LIM & Lima & Perú & (12.1 S, 77.02 W) & subtropical summer & 150 & 12.0 & 390 & -5 \\
    \hline
    LSC & La Serena & Chile & (29.90 S, 71.25 W) & subtropical summer & 28 & 9.3 & 380 & -3 \\
    \hline
    EZE & Buenos Aires & Argentina & (34.54 S, 58.44 W) & subtropical summer & 10 & 8.2 & 390 & -2 \\
    \hline
  \end{tabular}
  \end{center}
\end{table*}

\begin{figure}[ht]
  \centering
	\includegraphics[scale=0.65]{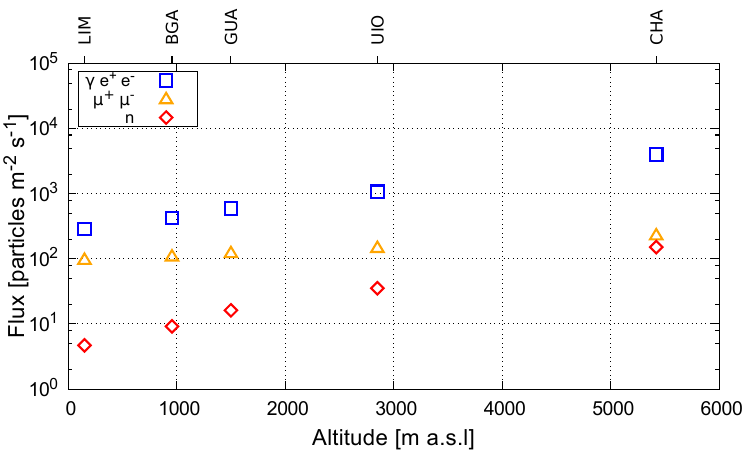}
	\includegraphics[scale=0.65]{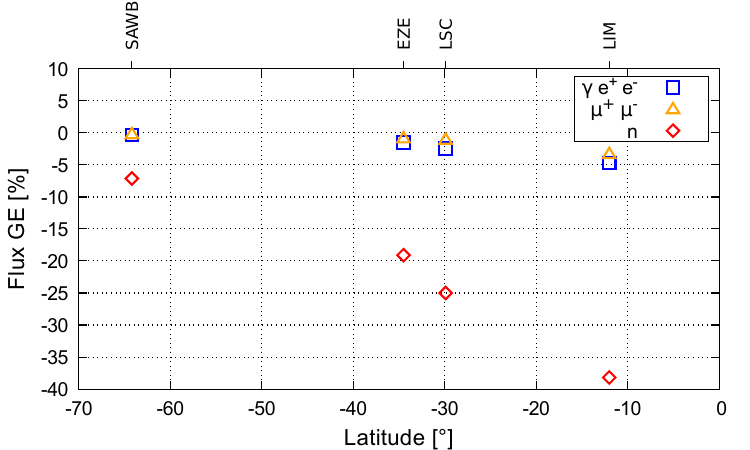}
	\caption{
        The flux ($\Xi$) of cosmic background radiation at the ground as a function of the altitude at five LAGO sites with very different altitudes but similar geomagnetic rigidities are shown in the upper panel for different components of the cascades: the electromagnetic component in blue squares ($\gamma$, $e^-$, $e^+$), the muonic component in yellow triangles ($\mu^-$, $\mu^+$) and neutrons as part of the hadronic component are shown in red diamonds ($n$). The altitude effect on the flux is visible. Instead, in the lower panel, the same flux is shown as a function of the geomagnetic rigidity for four LAGO sites with similar altitudes but very different values of the local rigidity cut-offs $R_\mathrm{c}$. While the effect on the total flux is not as significant as the altitude effect, the impact of the latitude on the neutron flux is quite significant.
	}\label{fig:components}
\end{figure}

This effect is also noticeable in Figure~\ref{fig:components}, which shows both effects for different EAS components: the atmospheric absorption changing with the altitude, and the EMF shielding depending on the latitude.
For example, the geomagnetic effect influences only the low energy primary particles, and so, the impact on the total secondary particle flux at ground level is more important on high altitude sites, where the impact of very low energy primaries is more significant.
However, the geomagnetic effect on the flux of secondary neutrons is dramatic, as can be seen in the lower panel of Figure~\ref{fig:components}; and this is one of the main reasons for using the neutron flux as an indicator to study the Solar activity from ground level observatories.
In this Figure we show a comparison in the flux of electromagnetic, muon and neutron components, at four sites with similar altitude but very different geomagnetic rigidities: Marambio (SAWB), Buenos Aires (EZE), La Serena (LSC), and Lima (LIM).
This considerable latitudinal dependence of the flux of secondary neutrons, combined with the capacity of WCDs to indirectly observe neutrons~\cite{sidelnik2019enhancing}, has a major impact on space weather studies, and strongly supports the installation of WCDs at sites with very low rigidity cut-off, such as the ones deployed by LAGO in the Antarctic continent.

\section{Water Cherenkov Detector Simulation}\label{sec:secwcd}

The third main stage of ARTI corresponds to a complete and detailed Geant4-based-simulation on a water Cherenkov detector~\cite{Agostinelli2003}.
As previously, the main result of the previous stage  --the resulting shower {\texttt{.shw}} file after considering the local geomagnetic field effects-- is used as the main input for this last stage.

A WCD consists of a closed, light-tight water container with optical detectors, typically PMTs or solid state photon counters, which are in close contact with the water volume so as to register the Cherenkov radiation.
The corresponding Cherenkov photons propagate inside the water volume and may be absorbed in the water or in the inner coating of the container or, instead, reach the optical detector producing a variable number of photo-electrons (pe).
These pe generate a current that is multiplied in the corresponding stages of the optical detector producing a pulse.
This small signal will be amplified and shaped by the detector electronics, to produce the so called pulse, i.e., the time evolution of the total charge deposited in the optical detector that it is measured at specific sampling rates.
Finally, if this pulse complies with a pre-established trigger conditions, it is registered, transmitted and stored for further processing, analysis and curation.

The duration and shape of the pulse registered will depend on several factors: 
\begin{itemize}
    \item the type and energy of the particle impinging the detector and its corresponding response;
    \item the water quality, 
    \item the detector geometry and age;
    \item the characteristics of the optical device;
    \item the sampling rate and even the components of the detector electronics
\end{itemize}
are some of the main factors needed to be taken into account by the simulation to represent the detector behaviour in the most possible accurate way.

The ARTI third stage incorporates different Geant4 macros and C++ codes so as to configure the detector characteristics and the type of processing that should be done with the corresponding simulated signals.
It also includes an injector macro that reads the secondaries {\texttt{.shw}} file and injects the particles in the water volume, and different Geant4 physics lists depending on the type and energy of the injected particle.
All the physical processes that need to be taken into account are considered: Cherenkov production in water, neutron scattering and neutron-proton reaction, muon and electronic captures, pair production, Compton scattering, photoelectric effect, Cherenkov propagation and absorption in the water, Cherenkov absorption and diffusive reflection in the inner coating, are some of the main processes included.
The simulation starts by placing all the secondary particles on a normalized area at the detector top located just above the WCD, and they are propagated and injected into the detector with their corresponding momenta $p_s$.
Depending on the particle type, the corresponding physical processes can take place and the propagation of relativistic charged particles also produce Cherenkov photons.
As soon as a Cherenkov photon of wavelength $\lambda_{\mathrm{Ch}}$ is created, a Metropolis-Hastings algorithm is used to determine if this photon will either produce an electronic signal at the optical detector or not.
The detection probability is calculated from the optical detector response using quantum efficiency reported by manufacturer, typically $\mathrm{QE}(\lambda{\mathrm{Ch}}) \lessapprox 25-30\%$.
This largely improves the computing efficiency, as only those Cherenkov photons that will eventually produce electronic signals are propagated inside the detector volume ($\lessapprox 5-10\%$).

\begin{figure}[ht]
\centering
  \includegraphics[scale=0.45]{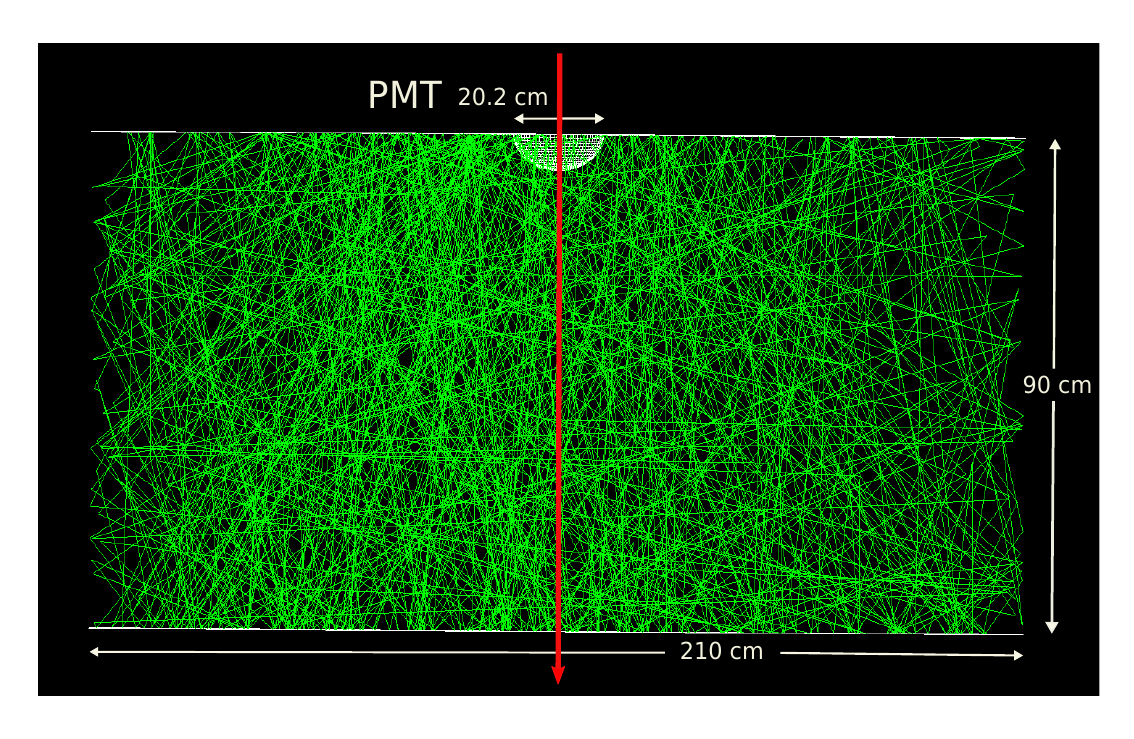}
  \includegraphics[scale=0.35]{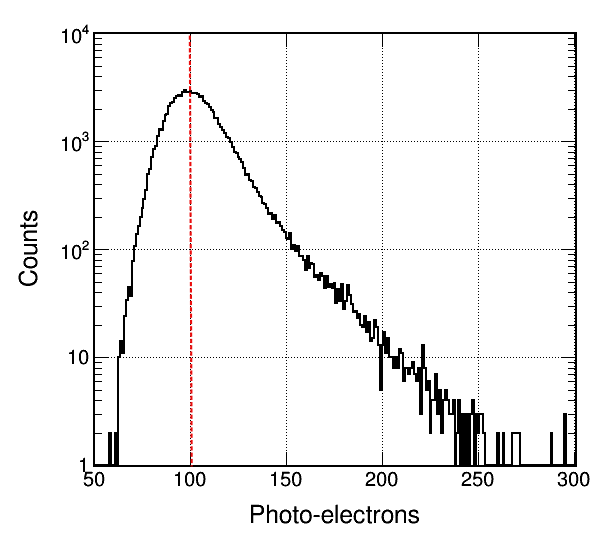}
	\caption{
		Top: A vertical and central muon of $3$\,GeV (red arrow) moves through the detector producing thousands of Cherenkov photons (green lines) that are propagated in the water volume and diffusely reflected in the Tyvek{\textregistered} coating. Eventually, these photons can be either absorbed or finally reach the PMT, producing a certain number of photo-electrons (pe).
        Bottom: distribution of the expected number of pe obtained due to the passage of $10^5$ central and vertical muons of $E_s=3$\,GeV, peaked at a value of $100$\,pe per VEM, that represents the main unit of calibration for this particular	simulated detector. By considering the muon stopping power in water at these energies, then $1$\,VEM$\simeq 100$\,pe$\simeq 180$\,MeV.
		}
    \label{fig:VEM}
\end{figure}

As an example of the ARTI capabilities, in this work we have evaluated the expected performance of the same generic LAGO WCD installed at the previously described sites.
LAGO WCDs are typically built with commercial plastic or stainless-steel cylindrical tanks filled with $1$\,m$^3$ to $4$\,m$^3$ of purified water and with an inner coating made of Tyvek{\textregistered}, a highly reflective and diffusive material at the ultraviolet spectrum region~\cite{filevich1999spectral,calderon2015geant4}.
A single 8"-9" PMT is located at the top centre of the tank, pointing downwards and in close contact with the water surface~\cite{Sidelnik2015}.
In this trial run, the simulated detector is a cylindrical tank ($h=0.90$\,m and $r=1.05$\,m), with the Tyvek coated inner surface, containing $3.12$\,m$^3$ of pure water, and an 8" Hamamatsu R5912 PMT\@.
After fixing the geometry and characteristics of the detector and its constituents, we use the same calibration procedure that as used at most astroparticle observatories using WCD~\cite{LAGO2015todos,bertou2006calibration,Galindo:2016P5}: signals are measured in VEMs, i.e., the total signal produced by the passage of vertical and central muons trough the water volume, as sketched in the upper panel of Figure~\ref{fig:VEM}.
Muons are typically used for the calibration of the deposited energy in WCD, and at the typical energy range of atmospheric muons ($2-6$\,GeV), their stopping power is nearly constant and close to the so-called MIP (minimum ionization particle), with an energy loss of $\sim 2$\,MeV\,cm$^2$\,g$^{-1}$, i.e., $2$\,MeV\,cm$^{-1}$ in water.
To estimate the signal value corresponding to $1$\,VEM, we inject in the detector vertical axis $10^5$ monoenergetic muons of $3$\,GeV moving in the $-z$ direction and obtaining the distribution shown in the lower panel of Figure~\ref{fig:VEM}.
The number of pe produced in the PMT is $60\lesssim \mathrm{VEM}_{\mathrm{pe}} \lesssim 250$, which peaked at $1 \mathrm{VEM}_{\mathrm{pe}} \simeq 100$\,pe, and thus $1$\,VEM$\simeq 100$\,pe$\simeq 180$\,MeV of deposited energy, $E_d$.

Once the VEM was obtained, the standard WCD calibration relies on identifying the main features of the so-called charge histogram, i.e., the histogram of the total charge by each of the secondary particles present in the flux $\Xi$ at the detector level~\cite{bertou2006calibration}.
While the exact characteristics of these features are essentially determined by the detector geometry and its response to the different EAS components, all the charge histograms share the same characteristics: a) a peak at low $E_d$ values corresponding to the detector response to the EM component and the trigger effect;
b) a second peak at larger $E_d$ values, the so called muon hump, product of the detector response to muons and corresponding to the typical charge produced by vertical muons, i.e., from the position of this hump in the measured histogram the value of the VEM can be obtained~\cite{aab2020studies}; and c) a change in the histogram slope at higher values of $E_d$ corresponding to the transition from single to multiple particles impinging in the detector~\cite{Asorey2015a}.

\begin{figure*}[ht]
\centering
	\includegraphics[scale=0.61]{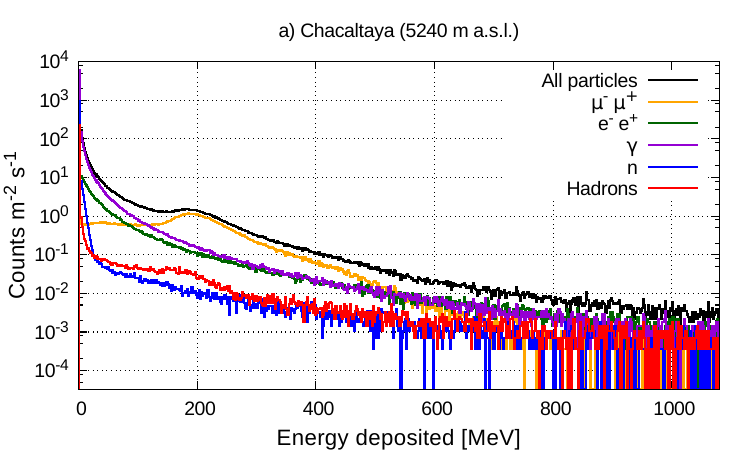}
	\includegraphics[scale=0.61]{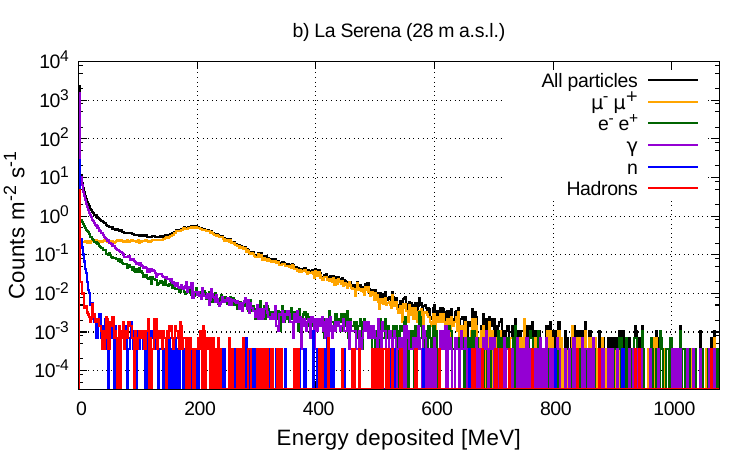}\\
	\includegraphics[scale=0.61]{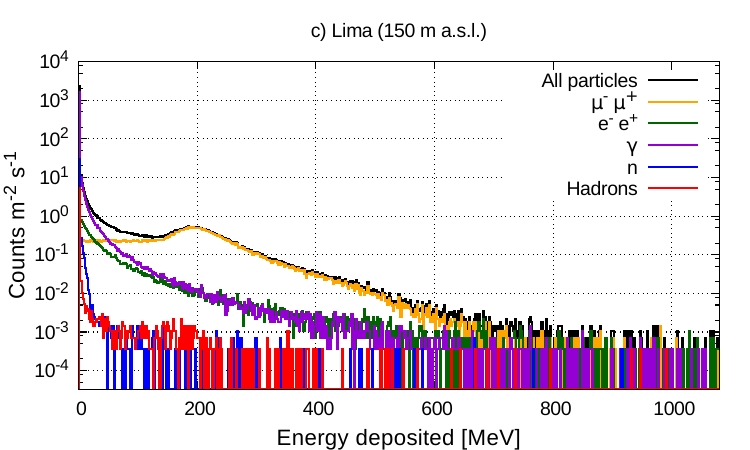}
	\includegraphics[scale=0.61]{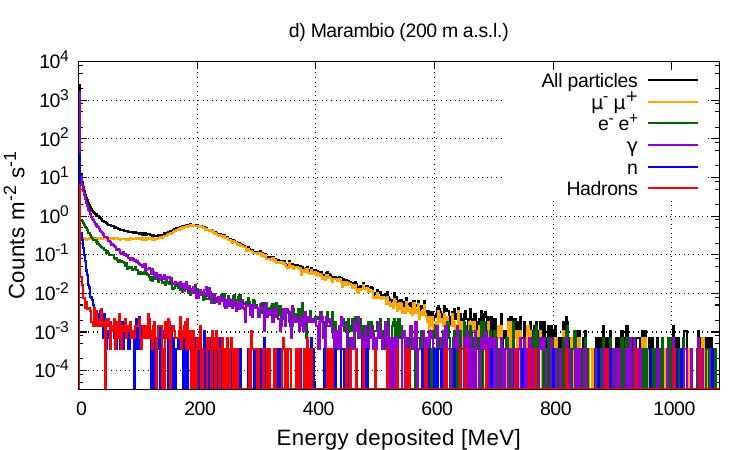}
	\caption{Charge histograms for four LAGO sites of the total signal (black line) and single component contributions (colour lines): Chacaltaya (a), La Serena (b), Lima (c) and Marambio (d). The histograms have been normalized to areas of $1$\,m$^2$ and exposure time of $1$\, second.}\label{fig:chargeHist}
\end{figure*}

This features can be observed in the simulated charge histograms obtained by ARTI and shown in figure~\ref{fig:chargeHist}.
These histograms were obtained by simulating the response of the same generic LAGO detector exposed to the total secondary flux $\Xi$ at the sites of Chacaltaya, La Serena, Lima and Marambio.
For all detectors, the deposited energy for the muon hump, ranges from $150$\,MeV\,$\lesssim$\,E$_d\lesssim$\,300\,MeV with slight differences between the sites, is peaked for all  cases at a value consistent with the VEM estimations, i.e., $1$\,VEM$\simeq 180$\,MeV\@.
The contribution of the different components of $\Xi$ to the charge histogram also depends on altitude; since at higher energy the fraction of muons is not as dominant as near sea level, where the EAS are totally developed and the EM component is more absorbed than the muon component in the atmosphere.

Of course, most of the low energy particles present in the total flux $\Xi$ are not be detected by a WCD and so the total signal flux $\Xi^{\mathrm{D}}$ will be lower that $\Xi$, as shown in Table~\ref{tab:energy_deposited}, where $\Xi^D$ is presented along with the percent reduction of about $55-60\%$ when compared with $\Xi$.
The table also shown the different contributions of each EAS component to the total deposited energy in the detector.

\begin{table}[ht]
	\caption{
        Expected signal flux, $\Xi^D$, in a single LAGO like WCD deployed at the ground level of the eight selected sites used as examples of the ARTI capabilities in this work.
        The percent relative difference of the expected signals to the expected flux of particles, i.e., $\Delta \Xi/\Xi = (\Xi^{\mathrm{D}}-\Xi)/\Xi \%$, and the deposited energy for the different EAS components of the are also shown: electromagnetic component E$_\mathrm{d}^{\mathrm{EM}}$, muon component E$_\mathrm{d}^{\mu}$, neutrons E$_\mathrm{d}^{\mathrm{n}}$, and the total deposited energy E$_\mathrm{d}^{\mathrm{All}}$.
        For the hadronic component, we only show the neutron contribution.
        The remaining hadronic contribution can be obtained by subtraction.
	}
	\begin{center}
			\begin{tabular}{|c|c|c|c|c|c|c|}
				\hline
                Site & $\Xi^D$ & $\Delta \Xi/\Xi$ & $E_d^{\mathrm{EM}}$ & $E_d^{\mu}$ & $E_d^{\mathrm{n}}$ & $E_d^{\mathrm{All}}$ \\
                \hline
                \hline
                CHA  & 1800 & -60 &  1.49 & 0.22 & 0.49 & 1.77 \\ \hline
                UIO  &  520 & -59 & 0.40 & 0.14 & 0.11 & 0.55\\ \hline
                GUA  &  310 & -57 & 0.22 & 0.11 & 0.05 & 0.34 \\ \hline
                BGA  &  230 & -57 & 0.16 & 0.10 & 0.03 & 0.26 \\ \hline
                SAWB &  180 & -58 & 0.11 & 0.10 & 0.02 & 0.21 \\ \hline
                LIM  &  170 & -56 & 0.11 & 0.09 & 0.02 & 0.20 \\ \hline
                LSC  &  170 & -55 & 0.11 & 0.09 & 0.01 & 0.20 \\ \hline
                EZE  &  170 & -56 & 0.11 & 0.09 & 0.01 & 0.20 \\ \hline
            \end{tabular}
    \end{center}\label{tab:energy_deposited}
\end{table}

\begin{figure}[ht]
  \centering
  \includegraphics[scale=0.675]{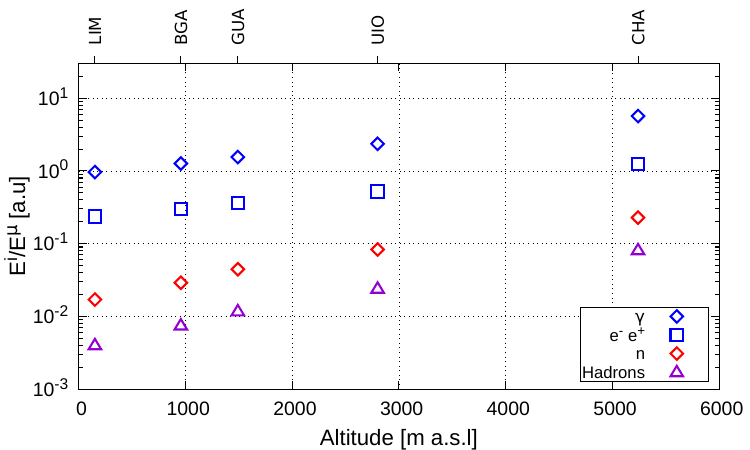}
	\caption{
        The component-to-muon ratio, i.e., the ratio between the deposited energy within the detector volume by a particular shower component (photons, blue diamonds; electrons, blue squares; neutrons, red diamonds; and hadrons, violet triangles) divided by the energy deposited by muons, is shown for five LAGO sites with similar geomagnetic rigidity but increasing altitude at Lima, Bucaramanga, Guatemala, Quito and Chacaltaya. The observed trend for this ratio reflects the expected evolution of the cascades with altitude, as the impact of the atmospheric absorption is much lower for muons than for other shower components. High altitude sites are optimal for studying photon-initiated showers, such as those produced by GRBs.
	}\label{fig:ratioMuon}
\end{figure}

Another parameter that is related with the EAS evolution is the relative fractions of the different components to the total muon content, since as the cascade evolves charged pions and kaons start to decay in to muons. When the shower $X_{\max}$ is behind, the electromagnetic component is being absorbed by the atmosphere.
Thus, it is to be expected that the muon component become more dominant as the altitude decreases, and so it's impact on the total deposited energy. This behaviour can be seen in Figure~\ref{fig:ratioMuon}, where the relative ratios in the deposited energy at the detector by muons and other components, i.e.  $E_{\mathrm{d}}^{\mathrm{i}}/E_{\mathrm{d}}^{\mathrm{\mu}}$; where $i$ represents the rest of the cascade component (photons, electrons, neutrons and other hadrons), are shown as a function of altitude, and as a possible observational trigger in our detectors for the ratio $\Xi^{\mathrm{i}}/\Xi^{\mathrm{\mu}}$.
This can be used to determine, e.g., the optimal altitude for detecting photon-initiated showers, as those produced during the sudden occurrence of a GRB~\cite{sarmiento2021latin}; or for space weather studies, as in this case we want to measure the Solar activity effects on the measured flux $\Xi^{\mathrm{D}}$.

\section{Conclusions}\label{sec:discussions}

In this work, we present a detailed description of the ARTI simulation toolkit, a publicly available and highly configurable framework designed to obtain in a semi-autonomous way, the detailed flux of the secondary particles and the corresponding signals at any type of water Cherenkov detector, produced by the interaction of the total flux of galactic cosmic rays with the atmosphere;
as well as, for example, the expected flux at ground produced by GRBs or other transient astrophysical phenomena.
Even more, this can also be done for time-evolving atmospheric and geomagnetic conditions at any place in the World, and including, e.g., the fast disturbances produced in the Earth's magnetic field by the Solar activity.

As an example of the ARTI capabilities, we calculated the signal flux expected at a simulated WCD virtually deployed at eight astroparticle observation sites in Latin America.
These sites were selected due to their distinctive characteristics in altitude (different atmospheric depths), and in latitude (different geomagnetic responses).
Throughout this work, the three main stages of the ARTI simulation were presented and as well as the description of the physical basis for all the calculations and assumptions.

In the first stage, based on user selections, ARTI performs the calculation of the total number of primaries to be injected into the chosen type of atmosphere and calls on the corresponding CORSIKA routines to simulate the atmospheric response to the primary flux.
Then, in the second stage, ARTI analyzes the first stage results to produce the expected flux of secondaries at ground level, $\Xi$, corrected by the real-time geomagnetic conditions.
Finally, during the third stage, the ARTI Geant4 macros are properly configured and used to simulate the expected response of a simulated water Cherenkov detector to the flux of secondaries.

ARTI can be easily obtained from the LAGO github repository~\cite{arti}, and can be prepared for running automatically at different types of computing facilities: from personal notebooks or desktops up to high performance computing clusters. More recently at cloud-based environments trough {\texttt{OneDataSim}}~\cite{rubiomontero2021eosc}, the ARTI implementation for the European Open Scientific Cloud (EOSC) and other federated and public clouds, which is publicly available at the EOSC Marketplace~\cite{marketplace}.

With the help of ARTI, we were able 
\begin{itemize}
    \item to characterize the expected response and sensitivity of LAGO or any other astroparticle observatory to the flux of cosmic rays in the galactic energy range~\cite{sidelnik2017lago,auger2020studies};
    \item to determine the sensitivity of LAGO at high altitude sites for the observation of steady gamma sources or astrophysical transients, such as the sudden occurrence of a gamma ray burst within the LAGO field of view~\cite{sarmiento2021latin};
    \item to study the impact of space weather phenomena from ground level by using water Cherenkov detectors~\cite{Asorey2015a,sidelnik2020simulation,Sarmiento2019modeling,rubiomontero2021eosc};
    \item to calculate the most statistically significant flux of high energy muons at underground laboratories, equivalent to one year of the expected primary flux at the site~\cite{rubiomontero2021eosc,perezbertoli2022estimation};
    \item to help in the assessment of active volcanoes risks in Latin America~\cite{pena2022muography,taboada2022meiga,vasquez2019simulated,vesgaramirez2021simulated};
    \item to design new safeguard radiation detectors for detecting the traffic of fissile materials~\cite{sidelnik2019enhancing,sidelnik2020neutron};
    \item to contribute to the detection of improvised explosive devise at warfare fields in Colombia~\cite{Vasquezramirez2021improvised};
    \item to estimate the expected radiation dose received by the crew during commercial flights;
    \item and even to determine the radiation exposure of equipment and people at Mars' surface during the incoming exploration missions.
\end{itemize}

\section{Acknowledgments}\label{sec:acknowledgment}

The LAGO Collaboration is very grateful to all the participating institutions and to the Pierre Auger Collaboration for their continuous support.
HA thanks Rafa Mayo for his warm welcome and continuous support during his stay at CIEMAT in Madrid, Spain.
The authors are grateful to Adrian José Pablo Sedoski (ITeDA), Alexander Martinez (UIS), Antonio Juan Rubio-Montero (CIEMAT), Angelines Alberto-Morilla (CIEMAT) and Alfonso Pardo-Diaz (CETA/CIEMAT) for their continuous support and fruitful computing discussions.
Some results presented in this paper were carried out using: a) the GridUIS-2 experimental test bed, being developed under the Universidad Industrial de Santander (SC3UIS) High Performance and Scientific Computing Center, development action with support from UIS Vicerrectoria de Investigación y Extension (VIE-UIS) and several UIS research groups as well as other funding bodies;
b) the Acme cluster, which is owned by CIEMAT and funded by the Spanish Ministry of Science and Innovation project CODEC-OSE (RTI2018-096006-B-I00) with FEDER funds as well as supported by the CYTED co-founded RICAP Network (517RT0529); and the Halley (UIS, Colombia) and the ITeDA (CNEA-CONICET-UNSAM, Argentina) local clusters.
LAN gratefully acknowledges the support of the Vicerrectoría de Investigación y Extensión from Universidad Industrial de Santander under project VIE2814.
HA and IS gratefully acknowledges the support from CNEA, CONICET and Agencia Nacional de Promoción de la Investigación, el Desarrollo Tecnológico y la Innovación (Agencia), for their financial support.
This work has been partially funded by the co-funded European Union’s Horizon 2020 research and innovation programme project ``European Open Science Cloud - Expanding Capacities by building Capabilities (EOSC-SYNERGY)'', under grant agreement No 857647.


\newpage

\appendix
\onecolumn

\section{ARTI pseudocode}\label{sec:pseudocode}

The following algorithm represents the three main stages that make up ARTI for flux simulations and EAS developments via CORSIKA, magnetic field correction via Magnetocosmic and detector simulation via Geant4.
\begin{algorithm}
    \SetKwInOut{Input}{Input}
    \SetKwInOut{Output}{Output}

    \underline{Simulation flux}\;
    \vspace{0.25cm}
    \Input{E = [1, 10$^6$] GeV; energy range\\
    Injected primary nuclei, Z = [1, 26] \\
	$\theta$ = [0, 90]; zenith angle range\\
	$\phi$ = [-180, 180]; azimuth angle range\\
	Altitude in m a.s.l. \\
	B$_x$, B$_z$; site's magnetic field\\
	atmospheric model of the site\\
	time;\\}
	\vspace{0.25cm}
    \Output{$\Xi$, particle flux at ground}
    \vspace{0.25cm}
    \Begin{
    $\Phi(E_p, Z, A, \Omega)$; Integrate cosmic-ray spectra\\
	Z, \#part(E) $\rightarrow$ built steering Corsika files\\
	run block, via Corsika software\\
	Analysis block; read and uncompress binary files\\
	}
      {
        return $\Xi$\;
      }

    \underline{Magnetic field correction}\;
    \vspace{0.25cm}
    \Input{$\Xi$, particle flux at ground\\
	Rm, magnetic rigidity\\
	IGRF, magnetic model\\}
    \Output{$\Xi_{corr}$, particle flux at ground corrected}
    \Begin{
    $R_\mathrm{C}(\phi, \theta)$, magnetic rigidity cutoff\\
    }
      {
        return $\Xi_{corr}$\;
      }
    \vspace{0.25cm}
    \underline{Water Cherenkov detector simulation}\;
    \vspace{0.25cm}
    \Input{$\Xi_{corr}$, geomagnetic corrected particle flux at ground\\
	D(r,h), detector's dimensions and characteristics\\
    W($\lambda$), water properties\\
	PMT, photo-multiplier's features\\
	}
	\Output{E$_{D}$, energy deposited}
	\vspace{0.25cm}
    \For{particles}
    {Interacting with water\\
    E$^{D}_{i}$, energy deposited by i-particle}
      {
        return E$_{D}$ \& charge histograms\;
      }
\vspace{0.5cm}
    \caption{ARTI is divided in three consecutive stages: flux calculations, atmospheric model and geomagnetic conditions to simulate EAS developments via Corsika; geomagnetic secular variations and disturbances via Magnetocosmic and their impact on the particle flux at ground; and the detector simulation via Geant4.}\label{alg:algorithm}
\end{algorithm}

\newpage

\section{ARTI Command Line Options Example}\label{sec:tutorial}

As described in the text, the simulation can be totally configured by selecting the corresponding options from the command line at launch time.
These include the flux time, the CORSIKA version, the observatory site, the number of process to use, the activation of the Cherenkov mode in the shower, etc.
All the user options override the default values included in ARTI\@.
As an example, here we show the currently available options for the main scripts of the first two ARTI stages: {\texttt{do\_sims.sh}} and {\texttt{do\_showers.sh}}.
All the ARTI scripts have their internal help that can be seen by using the {\texttt{-?}} modifier.

\begin{verbatim}
$ do_sims.sh -?

USAGE do_sims.sh:

Simulation parameters
  -w <working dir>                   : Working directory, where bin (run) files are located
  -p <project name>                  : Project name (suggested format: NAMEXX)
  -v <CORSIKA version>               : CORSIKA version
  -h <HE Int Model (EPOS|QGSII)>     : Define the high interaction model to be used
  -u <user name>                     : User Name.
  -j <procs>                         : Number of processors to use
Physical parameters
  -t <flux time>                     : Flux time (in seconds) for simulations
  -m <Low edge zenith angle>         : Low edge of zenith angle.
  -n <High edge zenith angle>        : High edge of zenith angle.
  -r <Low primary particle energy>   : Lower limit of the primary particle energy.
  -i <Upper primary particle energy> : Upper limit of the primary particle energy.
  -a <high energy ecuts>             : High energy cuts for ECUTS
  -y                                 : Select volumetric detector mode (default=flat array)
Site parameters
  -s <site>                          : Location (several options)
  -k <altitude, in cm>               : Fix altitude, even for predefined sites
  -c <atm_model>                     : Fix Atmospheric Model even for predefined sites.
  -o <BX>                            : Horizontal comp. of the Earth's mag. field.
  -q <BZ>                            : Vertical comp. of the Earth's mag. field.
  -b <rigidity cutoff>               : Rigidity cutoff; (if set value in GV = enabled).
  -g <Lat, Lon>                      : Obtain the current values of BX and BZ for a
                                       site located at (Lat,Lon,Alt). Unless the -s option
                                       is used, -k option is mandatory.
Modifiers
  -l                                 : Enables SLURM cluster compatibility (with sbatch).
  -e                                 : Enable CHERENKOV mode
  -d                                 : Enable DEBUG mode
  -x                                 : Enable other defaults (It doesn't prompt user
                                       for unset parameters)
  -?                                 : Shows this help and exit.
\end{verbatim}{}
\newpage

\begin{verbatim}
$ do_showers.sh -?

USAGE do_showers.sh:

  -o <origin directory>     : Origin dir, where the DAT files are located
  -r <ARTI directory>       : ARTI installation directory, generally pointed by $LAGO_ARTI
                              environment variable (default)
  -w <working directory>    : Working dir, where the analysis will be done (default is current
                              directory)
  -e <energy bins>          : Number of energy secondary bins (default: 20)
  -d <distance bins>        : Number of distance secondary bins (default: 20)
  -p <project base name>    : Base name for identification of S1 files (don't use spaces).
                              Default: odir basename
  -k <site altitude, in m>  : For curved mode (default), site altitude in m a.s.l. (mandatory)
  -s <type>                 : Filter secondaries by type: 1: EM, 2: MU, 3: HD
  -t <time>                 : Normalize energy distribution in particles/(m2 s bin), S=1 m2;
                              <t> = flux time (s).
  -m <bins per decade>      : Produce files with the energy distribution of the primary flux
                              per nuclei.
  -g <file> <col>           : Include geomagentic effects. Read the local rigidities from
                              column <col> of <file>.
  -j                        : Produce a batch file for parallel processing. Not compatible
                              with local (-l)
  -l                        : Enable parallel execution locally (N procs). Not compatible
                              with parallel (-j)
  -v                        : Enable verbosity. Will write .log processing files.
  -?                        : Shows this help and exit.
\end{verbatim}{}
\end{document}